\documentclass[12pt]{article}
\usepackage{epsfig}
\usepackage[centerlast,footnotesize]{caption2}
\begin{document}

\begin{center}
{\Large \bf Mass Spectra in the Doubly Symmetric Theory of
Infinite-Component Fields} \\

\vspace{0.5 cm}

\renewcommand{\thefootnote}{*}
L.M.Slad\footnote{E-mail: slad@theory.sinp.msu.ru} \\

\vspace{0.3 cm}

{\it D.V. Skobeltsyn Institute of Nuclear Physics,
Moscow State University, Moscow 119899, Russia}
\end{center}

\vspace{0.5 cm}

\centerline{\bf Abstract}

\begin{footnotesize}
We consider the problem of the characteristics of mass spectra in the doubly 
symmetric theory of fields transforming under the proper Lorentz group 
representations decomposable into an infinite direct sum of finite-dimensional 
irreducible representations. We show that there exists a range of free 
parameters of the theory where the mass spectra of fermions are quite 
satisfactory from the physical standpoint and correspond to the picture 
expected in the parton model of hadrons.
\end{footnotesize}

\vspace{0.5 cm}

\begin{small}
\begin{center}
{\large \bf 1. Introduction}
\end{center}

In [1], [2], the beginning was laid for investigating the relativistically 
invariant theory of fields transforming under representations of the proper 
Lorentz group $L^{\uparrow}_{+}$ that are decomposable into an infinite direct 
sum of finite-dimensional irreducible representations (we say that such fields 
belong to the ISFIR class).

The structure of relativistically invariant Lagrangians of any free fields was 
established and described by Gelfand and Yaglom [3], [4]; such Lagrangians have 
the form
\begin{equation}
{\cal L}_{0} = \frac{i}{2}[(\Psi, \Gamma^{\mu} \partial_{\mu} \Psi ) - 
(\partial_{\mu} \Psi, \Gamma^{\mu} \Psi )] - (\Psi, R\Psi ).
\end{equation}
For fields of the ISFIR class, the matrix operators $\Gamma^{\mu}$ involved in 
Lagrangian (1) contain an infinite number of arbitrary constants. Because of 
this, restriction to the theory of the ISFIR-class fields was performed in [1];
this restriction was achieved by imposing the condition that Lagrangian (1) is 
also invariant under the secondary-symmetry 
transformations\renewcommand{\thefootnote}{1)}\footnote{The notion of double 
symmetry, consisting of the primary and secondary symmetries, was introduced in
[5]. It can be considered a generalization of supersymmetry and the symmetry of
the Gell-Mann--Levi $\sigma$-model [6].}
\begin{equation}
\Psi '(x) = \exp (-iD^{\mu}\theta_{\mu})\Psi (x).
\end{equation}
In formula (2), the parameters $\theta_{\mu}$ are polar or axial four-vectors 
of the orthochronous Lorentz group $L^{\uparrow}$, and the operators $D^{\mu}$
have a matrix realization. The chosen requirement leads to selecting a 
countable set of versions of the theory, each of which is characterized by a 
definite representation of the $L^{\uparrow}_{+}$ group, by single-valued (up 
to a normalization constant) four-vector operators $\Gamma^{\mu}$ and 
$D^{\mu}$, and by the operator $R$ being proportional to the unit operator $E$.

To avoid infinite degeneration with respect to spin in the mass spectrum for 
the Lorentz group extension performed in [1] (such a degeneration is to be 
expected in accordance with the Coleman--Mandula theorem [7]), spontaneous 
breaking of the secondary symmetry was postulated in [2]: it was assumed that
scalar components (with respect to the $L^{\uparrow}$ group) of one or several 
bosonic infinite-component fields of the ISFIR class have nonzero vacuum 
expectations $\lambda^{i}$ (the index $i$ can, e.g., denote a chosen component 
of some representation of the internal symmetry group $SU(3)$). Assuming that 
the operator $R$ specifying the mass term in the Lagrangian of the free 
fermion field is entirely caused by spontaneous breaking of the secondary 
symmetry, we then have
\begin{equation}
R = \sum_{i}  \lambda^{i} Q^{(0,1)00}_{i}.
\end{equation}
The operators $Q^{(0,1)00}_{i}$ occurring in relation (3) originate from the 
Lagrangian of the fermion-boson interaction
\begin{equation}
{\cal L}_{\rm int} = \sum_{i',\tau,l,m}  
(\psi (x), Q^{\tau lm}_{i'} \varphi_{\tau lm}^{i'} (x) \psi (x)),
\end{equation}
where $\psi (x)$ is the fermionic field, $\varphi_{\tau lm}^{i'} (x)$ is the 
component of the bosonic field characterized by the finite-dimensional 
irreducible representation  $\tau = (l_{0}, l_{1})$ of $L^{\uparrow}_{+}$, by 
the spin $l$ ($l = |l_{0}|, |l_{0}|+1, \ldots ,|l_{1}|-1$), and by its 
projection $m$ on the third axis, and $Q^{\tau lm}_{i'} \equiv 
Q^{(l_{0}, l_{1})lm}_{i'}$ are matrix operators. The index $i'$ in formula (4) 
can include some index $i$ from relation (3) and, in addition, information 
about transformation properties of the corresponding bosonic field under the 
spatial reflection. The problem of the existence of a nontrivial doubly 
symmetric Lagrangian (4) and of matrix elements of the operator 
$Q^{(0,1)00}_{i}$  was solved in [2].

Sharing the multiply expressed belief [8] that monolocal infinite-component 
fields can effectively represent composite particles and using the results in 
[1] and [2], we study the possibility of describing hadrons and their 
interactions using the doubly symmetric theory of fields of the ISFIR class. 
For this possibility to have any chance of being realistic, the theory under 
consideration must satisfy at least the following two requirements. First, it 
must ensure the existence of physically acceptable mass spectra. Second, the 
elastic electromagnetic interaction of the ground fermionic state of the theory
must be described by form factors similar to the nucleonic ones. The first 
requirement is addressed in this paper. 

We recall that just the mass spectra were the main stumbling block in the 
previous attempts at a consistent relativistic description of particles with an
infinite number of degrees of freedom, realized as different states of the same
field, because in all the cases considered, the spectra had an accumulation 
point at zero. This conclusion regarding the mass spectra was drawn, in 
particular, by Ginzburg and Tamm [9], Yukawa [10], Shirokov [11], and Markov 
[12] in their analyses of certain particular bilocal equations and also by 
Gelfand and Yaglom [3] and Komar et al. [13] with regard to the general linear
relativistically invariant equations based on representations of the group
$L^{\uparrow}_{+}$ that are decomposable into a finite direct sum of 
infinite-dimensional irreducible representations.

The result obtained in this work should therefore appear all the more 
important: the doubly symmetric theory of infinite-component fields of the 
ISFIR class has mass spectra that qualitatively reproduce the characteristic 
features of the experimental physics of hadrons and the parton bag model. We 
give its derivation and detailed formulation only for fermionic fields in the 
versions of the theory with one free parameter. Versions of the theory with 
three, five, and more free parameters with sufficiently high positions of the 
mass-spectrum levels (relative to the ground level) with any given precision 
reduce to the versions with a single parameter. Some of the mass-spectrum 
characteristics that we obtain occur as consequences of exact analytic 
calculations, and others are the result of applying numerical methods.

After solving the main problem, we give some attention in this paper to the 
problem of comparing the theoretical mass spectrum with the levels of hadron 
resonances. We then note a nontrivial feature of the ISFIR-class fields that a 
state of such a field can be manifested experimentally as a group of several 
nearby resonances with different spins. We indicate a possible identification 
of such groups in the case of nucleon resonances (see the table below). The 
confidence level of this identification cannot be established otherwise than by
using an extensive analysis involving, first, the expected correspondence 
between the theoretical and experimental nucleon and $\pi$-meson levels, 
second, the wave functions found for each of the theoretical states involved in
the analysis, and, third, comparison of the results of theoretical (based on 
Lagrangian (4)) and experimental partial-wave analysis of the production 
amplitude of $\pi N$-states in the ranges of the observed nucleon resonances.

\begin{center}
{\large \bf 2. Equations and conditions for state vectors of particles 
described by ISFIR-class fields}
\end{center}

The subject of our analysis is the linear relativistically invariant equations 
for a free fermionic field of the ISFIR class,
\begin{equation}
(\Gamma^{\mu} \partial_{\mu} + i R) \psi (x) = 0,
\end{equation}
corresponding to Lagrangians (1). For a field corresponding to a plane wave 
with a zero spatial wave vector $\psi (x) = \exp (-iMt) \psi_{M0}$, Eq. (5) 
becomes
\begin{equation}
(M \Gamma^{0} - R) \psi_{M0} = 0.
\end{equation}
For each allowed value of the spin and its third projection, this equation 
becomes one or several recursive relations for the components of an 
infinite-component field $\psi_{M0}$. For any value of $M$, these relations 
themselves give one or several linearly independent solutions up to a numerical
factor. 

If Eqs. (6) were a closed mathematical problem, then it would most likely be 
complemented by the normalizability condition for its solutions. But in 
particle physics, we are interested not so much in the fields as in the 
amplitudes of the processes that are expressed through currents of the form 
$(\psi_{M0}, O \psi_{Mp})$ in the theory under consideration. The field vector 
$\psi_{Mp}$ is obtained from the vector $\psi_{M0}$ via a transition from one 
reference frame to another such that the wave four-vector $\{ M,0,0,0 \}$ 
transforms into the four-vector $\{ E,0,0,p \}$. The role of the operator $O$ 
can be played by any of the matrix operators $Q^{\tau lm}_{i'}$, 
$\Gamma^{\mu}$, $D^{\mu}$, or $R$ in the theory. The currents 
$(\psi_{M0}, O \psi_{Mp})$ are in turn expressed through infinite series whose 
terms are quadratic in the components of the vectors $\psi_{M0}$ and linear in 
the matrix elements of finite transformations of the proper Lorentz group. 
Therefore, instead of the normalizability condition for solutions of Eq. (6), 
we introduce a more restrictive condition that the relevant series converge, 
which we call the finite-amplitude condition. For any of the operators $O$ 
above, the matrix elements relating the irreducible representations 
$\tau = (\pm 1/2, l_{1})$ and $\tau' = ( 1/2, l_{1}+n')$ or 
$\tau'' = (-1/2, l_{1}+n'')$, where $n'$ and $n''$ are integers, have the
asymptotic behavior as $l_{1} \rightarrow +\infty$ of the form $l_{1}^{\beta}$,
where $\beta$ is some constant. In view of this and of the asymptotic form of 
the field vector components $\psi_{M0}$ and $\psi_{Mp}$ as 
$l_{1} \rightarrow +\infty$, which is given in this work, we can verify that 
for any fixed value of $M$, either all the series corresponding to the above 
currents converge or all of them diverge. We therefore formulate the 
finite-amplitude condition for any value of $p$ as the relation
\begin{equation}
|(\psi_{M0}, R \psi_{Mp})| < +\infty
\end{equation}
or
\begin{equation}
(\psi_{M'0}, R \psi_{Mp}) = a(p) \delta (M'-M),
\end{equation}
where $a(p)$ is a nonzero number. Those and only those field vectors 
$\psi_{M0}$ that satisfy Eq. (6) and relations (7) or (8) are called state 
vectors of a mass-$M$ particle; this particle is a point in the respective 
discrete or continuum part of the mass spectrum.

Using formula (19) below and the appropriate analogue of formula (18), we can 
easily verify that if relation (8) is satisfied for some value $p = p_{0}$, 
then it cannot be satisfied for $p > p_{0}$. Therefore, in the theory of 
ISFIR-class fields under consideration, the mass spectrum cannot have a 
continuum part.

For $p = 0$, finite-amplitude condition (7) or (8) becomes the normalizability 
condition for solutions of Eq. (6). We note that the finite-amplitude condition
and the normalizability condition for solutions lead to the same mass spectra 
in all versions of the theory considered in Secs. 4--6 below. In some versions
considered in Sec. 3, there is a nonempty discrete mass spectrum if the 
normalizability condition for solutions of Eq. (6) is satisfied, whereas the 
finite-amplitude requirement leads to an empty mass spectrum.

We now give the needed results in [1] and [2] concerning the doubly symmetric 
theory of ISFIR-class fields with the spontaneously broken secondary symmetry 
considered here.

The $L^{\uparrow}_{+}$-group representation $S$, under which the fields
transforms, must coincide with one of the infinite-dimensional representations 
$S^{k_{1}}$ ($k_{1}$ is a half-integer, $k_{1} \geq 3/2$),
\begin{equation}
S^{k_{1}} = \sum^{+\infty}_{n_{1}=0} 
\sum^{k_{1}-\frac{3}{2}}_{n_{0}=-k_{1}+\frac{1}{2}}
\oplus (\frac{1}{2}+n_{0}, k_{1}+n_{1}).
\end{equation}

In the representation space of $S^{k_{1}}$, the operator $\Gamma^{0}$ is given 
by
$$\Gamma^{0} \xi_{(l_{0}, l_{1}) lm} = c_{0} D(l_{1}) 
\sqrt{\frac{(l-l_{0})(l+l_{0}+1)(k_{1}-l_{0}-1)(k_{1}+l_{0})}
{(l_{1}-l_{0})(l_{1}-l_{0}-1) (l_{1}+l_{0}) (l_{1}+l_{0}+1)}}
\xi_{(l_{0}+1, l_{1}) lm}$$
$$+ c_{0} D(l_{1}) 
\sqrt{\frac{(l-l_{0}+1)(l+l_{0})(k_{1}-l_{0})(k_{1}+l_{0}-1)}
{(l_{1}-l_{0}+1)(l_{1}-l_{0}) (l_{1}+l_{0}-1) (l_{1}+l_{0})}}
\xi_{(l_{0}-1, l_{1}) lm}$$
$$- c_{0} D(l_{0})
\sqrt{\frac{(l_{1}-l)(l_{1}+l+1)(l_{1}-k_{1}+1)(l_{1}+k_{1})}
{(l_{1}-l_{0})(l_{1}-l_{0}+1) (l_{1}+l_{0}) (l_{1}+l_{0}+1)}}
\xi_{(l_{0},l_{1}+1) lm}$$
\begin{equation}
- c_{0} D(l_{0})
\sqrt{\frac{(l_{1}-l-1)(l_{1}+l)(l_{1}-k_{1})(l_{1}+k_{1}-1)}
{(l_{1}-l_{0}-1)(l_{1}-l_{0}) (l_{1}+l_{0}-1) (l_{1}+l_{0})}}
\xi_{(l_{0},l_{1}-1) lm},
\end{equation}
where $\xi_{(l_{0}, l_{1}) lm}$ is a vector of the canonical basis in the space
of the irreducible representation, $\tau = (l_{0},l_{1}) \in S^{k_{1}}$, and 
$c_{0}$ is an arbitrary real constant. The function $D(j)$ of a half-integer 
argument $j$ is given by the formula
\begin{equation}
D(j) = 1
\end{equation}
if the secondary symmetry of the theory is generated by a polar four-vector 
representation of $L^{\uparrow}$ (Corollary 1 in [1]) and by the formula
\begin{equation}
D(j) = (-1)^{j-\frac{1}{2}} j
\end{equation}
if the secondary symmetry of the theory is generated by an axial four-vector 
representation of $L^{\uparrow}$ (Corollary 2 (A. 2) in [1]).

For the operator $R$ in the representation space $S^{k_{1}}$ and for both 
versions of the secondary symmetry in the considered theory, the same relations
hold,
\begin{equation}
R \xi_{(l_{0},l_{1})lm} = r(l_{0},l_{1}) \xi_{(l_{0},l_{1})lm} =
\left[ \sum_{i} \lambda^{i} q_{i}(l_{0},l_{1}) \right] \xi_{(l_{0},l_{1})lm},
\end{equation}
\begin{equation}
q_{i}(-l_{0},l_{1}) = q_{i}(l_{0},l_{1}),
\end{equation}
$$(k_{1}-l_{0}-1)(k_{1}+l_{0})q_{i}(l_{0}+1,l_{1})
+ (k_{1}-l_{0})(k_{1}+l_{0}-1)q_{i}(l_{0}-1,l_{1})$$
$$- (k_{1}-l_{1}-1)(k_{1}+l_{1})q_{i}(l_{0},l_{1}+1)
- (k_{1}-l_{1})(k_{1}+l_{1}-1)q_{i}(l_{0},l_{1}-1) =$$
\begin{equation}
= z_{i}(l_{1}-l_{0})(l_{1}+l_{0})q_{i}(l_{0},l_{1})
\end{equation}
with $z_{i} = 2-H_{i}^{\rm B}/H^{\rm F}$. The quantities $H_{i}^{\rm B}$ and 
$H^{\rm F}$ are normalization constants of the vector operators $D^{\mu}$ (with
$D^{\mu}D_{\mu} = HE$) entering transformations (2) of the respective bosonic 
and fermionic fields $\varphi^{i} (x)$ and $\psi (x)$. They are independent of 
each other. For some fixed value of $H^{\rm F}$, the quantity $H_{i}^{\rm B}$ 
and hence the parameter $z_{i}$ can take any values. Indeed, if secondary 
symmetry transformation (2) of the bosonic field $\varphi^{i} (x)$ is 
nontrivial ($D^{\mu} \neq 0$) and this field is complex (just such fields were 
considered in [1] and [2]), then $H_{i}^{\rm B} > 0$. If transformation (2) is 
nontrivial and the bosonic field is real, then $H_{i}^{\rm B} < 0$. But if 
transformation (2) of the bosonic field is trivial, then $z_{i} =2$, the 
operator $Q^{(0,1)00}_{i}$ is proportional to the unit operator $E$, and the 
existence of a nonzero vacuum expectation of this field does not affect the 
secondary symmetry.

In what follows, we only deal with fermionic fields transforming under the 
"lowest" of the proper Lorentz group representations $S^{k_{1}}$ in (9), 
namely, the representation $S^{3/2}$, whose decomposition involves all 
finite-dimensional irreducible representations that contain spin $1/2$ and only
such representations. A solution of recursive relation (15) for this 
representation ($k_{1} = 3/2$) was expressed in [2] through the Gegenbauer 
polynomials and through hypergeometric series. It turns out that this solution 
can also be expressed through the elementary functions,
\begin{equation}
q_{i}\left( -\frac{1}{2},l_{1}\right) = q_{i}\left( \frac{1}{2},l_{1}\right)  = 
2q_{i0}\frac{u_{i}^{N}(u_{i}N+N+1)-w_{i}^{N}(w_{i}N+N+1)}{N(N+1)(u_{i}-w_{i})
(2+u_{i}+w_{i})},
\end{equation}
where $N = l_{1}-1/2$, $u_{i}=(z_{i}+\sqrt{z_{i}^{2}-4})/2$, and 
$w_{i}= (z_{i}-\sqrt{z_{i}^{2}-4})/2$.

In accordance with formulas (10) and (13), the operators $\Gamma^{0}$ and $R$ 
are diagonal in the spin index $l$ and in the spin projection index $m$, and 
their matrix elements are independent of $m$. Therefore, each vector 
$\psi_{M0}$ satisfying Eq. (6) can be assigned certain values of spin and its 
projection. Components of this vector can be taken independent of the value of 
$m$. Linearly independent solutions of Eq. (6) can be chosen such that they 
have a definite $P$-parity. We recall that $P \xi_{(\pm 1/2,l_{1})lm} = 
(-1)^{l-1/2} \xi_{(\mp 1/2,l_{1})lm}$. If a vector 
$\sum_{l_{1}} [\chi (l_{1})\xi_{(-1/2,l_{1})lm} + 
\chi (l_{1}) \xi_{(1/2,l_{1})lm}]$ with the $P$-parity $(-1)^{l-1/2}$ satisfies 
condition (7) and Eq. (6) with $M=M_{0}$, then the vector    
$\sum_{l_{1}} [-\chi (l_{1}) \xi_{(-1/2,l_{1})lm} + 
\chi (l_{1}) \xi_{(1/2,l_{1})lm}]$, which has the $P$-parity $(-1)^{l+1/2}$, 
also satisfies condition (7) and Eq. (6) but with $M=-M_{0}$.

In the representation space $S^{3/2}$ of the $L^{\uparrow}_{+}$ group, Eq. (6) 
for the field vector components 
$(\psi_{M0})_{(\pm 1/2,l_{1})lm} \equiv \chi_{lm}(l_{1})$ with the spatial 
parity $(-1)^{l-1/2}$ becomes
$$D\left( \frac{1}{2} \right) \frac{\sqrt{(l_{1}-l)(l_{1}+l+1)}}{2l_{1}+1} 
\chi_{lm} (l_{1}+1)
+D\left( \frac{1}{2} \right) \frac{\sqrt{(l_{1}-l-1)(l_{1}+l)}}{2l_{1}-1} 
\chi_{lm} (l_{1}-1 )$$
\begin{equation}
- \left[ \frac{D(l_{1})(2l+1)}{4l_{1}^{2}-1}
- \frac{1}{2Mc_{0}} r(l_{1}) \right] \chi_{lm} (l_{1}) = 0,
\end{equation}
where $l_{1} \geq l$ and $r(l_{1}) \equiv r(\pm 1/2,l_{1})$.

We write the relativistically invariant bilinear form of relation (7) via 
components of the vectors $\chi_{lm}(l_{1})$ and via matrix elements of finite 
transformations of the proper Lorentz group,
\begin{equation}
(\psi_{M0}, R \psi_{Mp}) =
\sum_{l_{0}=-\frac{1}{2}}^{\frac{1}{2}} \sum_{l_{1}=l+1}^{+\infty} 
(-1)^{l-\frac{1}{2}} \chi^{*}_{lm} (l_{1}) r(l_{1}) 
\{ [\exp(\alpha I^{03})]_{(l_{0},l_{1})lm,(l_{0},l_{1})lm} \} \chi_{lm}(l_{1}),
\end{equation}
where $\tanh \alpha = p/\sqrt{M^{2}c^{2}+p^{2}}$ and $I^{03}$ is the 
infinitesimal operator of the group $L^{\uparrow}_{+}$. 

The explicit form of matrix elements of finite transformations of the 
$L^{\uparrow}_{+}$ group for infinite-dimensional unitary irreducible 
representations, denoted by a pair of numbers $(l_{0}, \nu)$, was found in 
[14]. The argument and the results in that work also hold for the 
finite-dimensional irreducible representations that interest us here. To 
preserve the precise meaning of the notation that we use, we must set 
$\nu = il_{1}$. In considering the convergence problem for the series of form 
(18), we need only know the asymptotic behavior of the relevant matrix elements
as $l_{1} \rightarrow +\infty$. We have
\begin{equation}
[\exp(\alpha I^{03})]_{(l_{0},l_{1})lm,(l_{0},l_{1})lm} = 
T_{0} \frac{\exp (\alpha l_{1})}{l_{1}} (1+{\cal O}(l_{1}^{-1})),
\end{equation}
where the quantity $T_{0}$ is independent of $l_{1}$.

In what follows, we discuss the mass spectra in two versions of spontaneous 
secondary symmetry breaking: caused by one bosonic field of the ISFIR class 
(in which case the index $i$ is to be omitted everywhere) or caused by two 
bosonic fields. In the first version, we separately consider three essentially
different ranges of the $z$ parameter values: $(-\infty , -2]$, $(-2, 2)$, and 
$(2, +\infty )$. In the second version, attention is given only to the range 
$z_{1} \in (2, +\infty)$ and $z_{2} \in (-2, 2)$. 

\begin{center}
{\large \bf 3. Empty mass spectrum in the parameter range $z \in (-2,2)$}
\end{center}

In the case where the secondary symmetry of the theory is spontaneously broken,
we cannot find solutions of Eq. (17) in the form of elementary or special 
functions, finite or infinite series. We also fail to find analytic formulas 
for mass spectra of the theory. But we can derive a number of conclusions 
regarding the mass spectra based on the asymptotic behavior of certain 
quantities.

Let $z \in (-2, 2)$. Using formula (16), we then have
\begin{equation}
r(l_{1}) = r_{0} \frac{\sin \zeta l_{1}}{l_{1}} (1+{\cal O}(l_{1}^{-1}))
\end{equation}
as $l_{1} \rightarrow +\infty$, where $\zeta \in (0, \pi )$ and $r_{0}$ is a 
constant. From this and Eq. (17), we obtain
\begin{equation}
\chi_{lm} (l_{1}) = A_{0} (-1)^{\left[ \frac{l_{1}}{2} \right]}
l_{1}^{s} ( 1+K(l_{1})+{\cal O}(l_{1}^{-2})) 
+ B_{0} (-1)^{\left[ \frac{l_{1}+1}{2} \right]}
l_{1}^{-s} ( 1-K(l_{1})+{\cal O}(l_{1}^{-2}))
\end{equation}
as $l_{1} \rightarrow +\infty$, where
\begin{equation}
s = (2l+1)\left( 1-D\left( \frac{1}{2} \right) \right), \hspace{0.5 cm}
K(l_{1}) = (-1)^{l_{1}+\frac{1}{2}} 
\frac{r_{0}}{2Mc_{0} D\left( \frac{1}{2} \right) \sin\zeta} 
\cdot \frac{\cos\zeta l_{1}}{l_{1}},
\end{equation}
$[a]$ is the integer part of the number $a$, and the quantities $A_{0}$ and 
$B_{0}$ are independent of $l_{1}$. 

Based on this asymptotic formula, it is easy to establish that for 
$\alpha \neq 0$, the terms of series (18) grow as $l_{1} \rightarrow +\infty$ 
independently of whether $A_{0}$ is equal to zero for certain values of $M$. 
Therefore, if $z \in (-2, 2)$, then a solution of Eq. (17) cannot satisfy 
finite-amplitude condition (7) at any value of $M$, and the mass spectrum is 
hence empty.

\begin{center}
{\large \bf 4. Characteristics of the mass spectra in the parameter range 
$z \in (-\infty, -2]$}
\end{center}

Because the inequalities $w < -1$ and $-1 < u < 0$ hold in the range 
$z \in (-\infty, -2)$, it follows from (16) and (17) that as 
$l_{1} \rightarrow +\infty$,
\begin{equation}
r(l_{1}) = r_{0} \frac{w^{l_{1}+\frac{1}{2}}}{l_{1}+\frac{1}{2}} 
(1+{\cal O}(l_{1}^{-1})),
\end{equation}
\begin{equation}
\chi_{lm}(l_{1}) = A_{0} G(l_{1}) (1+{\cal O}(l_{1}^{-1})) 
+ B_{0} G^{-1}(l_{1}) (1+{\cal O}(l_{1}^{-1})),
\end{equation}
where
\begin{equation}
G(l_{1}) = \frac{v^{l_{1}-\frac{1}{2}}w^{\frac{4l_{1}^{2}-1}{8}}}
{(l_{1}-\frac{1}{2})!},
\hspace{0.5 cm} v = -\frac{r_{0}}{Mc_{0}D(\frac{1}{2})}.
\end{equation}

Obviously, if $A_{0}$ is nonzero for some values of $M$, then condition (7) 
cannot be satisfied. But if $A_{0}=0$ for some value of $M$, then the terms of 
series (18) have the asymptotic form of the order of
$(l_{1}+1/2)!(l_{1}-1/2)!v^{-2l_{1}}u^{(4l_{1}^{2}-1)/4}\exp (\alpha l_{1})$ as
$l_{1} \rightarrow +\infty$. The ratio of such a term of the series to the 
preceding term is equal to zero in the limit $l_{1} \rightarrow +\infty$. 
Therefore, the relevant series (18) converges for the discussed value of $M$, 
which is equivalent to condition (7) being satisfied.

Therefore, for all values of the parameter $z$ in the range 
$z \in (-\infty, -2)$ and for both versions of the theory expressed by 
relations (11) and (12), the mass spectrum is discrete whenever it is nonempty.

A similar statement also holds for $z=-2$. This is easy to verify taking into 
account that in this case, $r(l_{1}) = r_{0}(-1)^{l_{1}+1/2}l_{1}$ and an 
analogue of relation (24) holds with 
$G(l_{1}) = v^{l_{1}-\frac{1}{2}}(-1)^{(4l_{1}^{2}-1)/8} (l_{1}-1/2)!$.

We now prove that in the range $z \in (-\infty, -2]$, the set of all masses of 
the theory is bounded from below by a positive number. For this, it suffices to
find a number $\mu_{0} > 0$ such that for the set of values of $M$ satisfying 
the restriction $|M| \leq \mu_{0}$, the field components lm(l1) are not 
arbitrarily small for sufficiently large values of $l_{1}$.

Using formula (16), we verify that in the range $z \leq -2$, the quantity 
$|r(l_{1})|$ 
increases monotonically as $l_{1}$ increases and the function $r(l_{1})$ of a 
half-integer argument has alternating signs, $r(l_{1}+1)/r(l_{1}) < 0$. Let
$\mu_{1}=|r(3/2)/c_{0}|$ and $|M| \leq \mu_{1}/2$. Then for both versions of 
the function $D(j)$ (Eqs. (11) and (12)) and for all the allowed values of $l$ 
and $l_{1}$, the quantity $D(l_{1})(l+1/2)/(l_{1}^{2}-1/4)-r(l_{1})/Mc_{0}$ is 
greater than one in absolute value and changes its sign as $l_{1}$ changes by 
$1$. Together with Eq. (17), this gives the desired inequality 
$|\chi_{lm}(l_{1}+1)| > |\chi_{lm}(l_{1})|$.

\begin{figure}[ht]
\vspace{-1.2cm}
\centering \includegraphics[width=14.0cm]{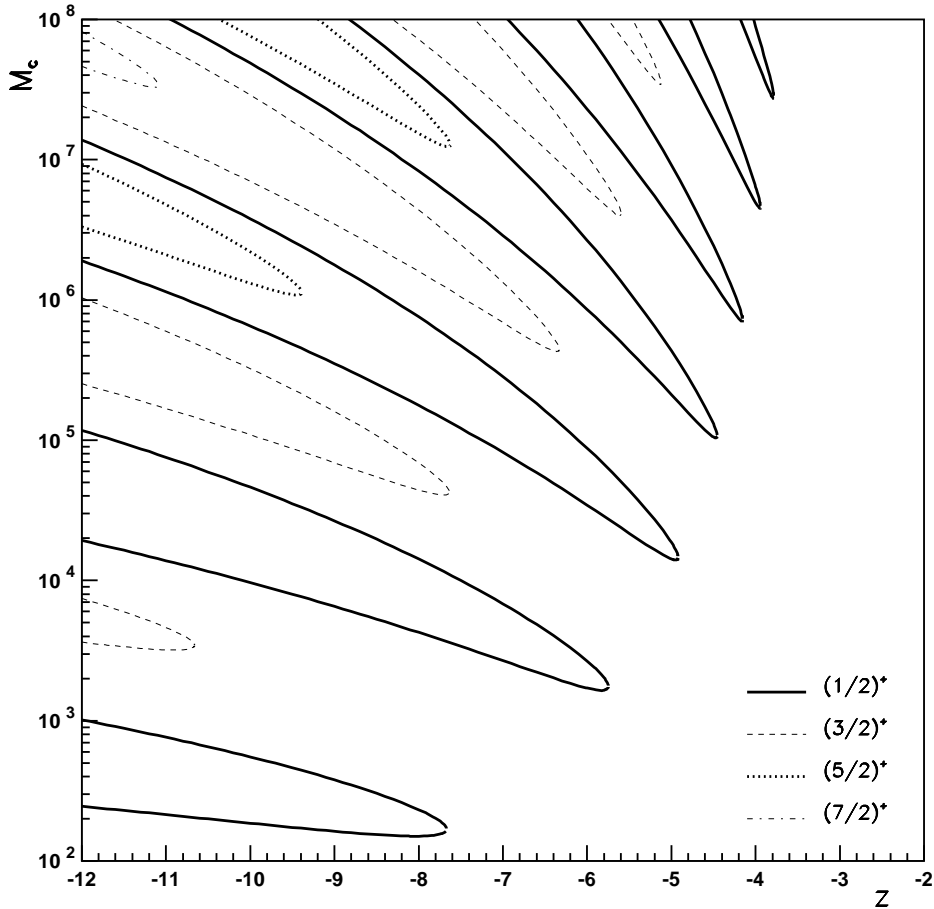}
\vspace*{-0.9cm}\\
\caption{Dependence of the mass levels on the parameter $z$ for $z < -2$ in the
theory with double symmetry generated by the polar four-vector representation 
of the orthochronous Lorentz group.}
\end{figure}

For convenience in what follows, we use one or another relation between the 
normalization constants, thus obtaining different mass units. If 
$r(3/2)/(2c_{0}D(1/2)) = \pm 1$ (the plus and minus signs refer to the 
respective relations (11) and (12), chosen such that the lower levels of the 
spin-$1/2$ particle have the spatial parity $+1$), we introduce the notation 
$M_{c}= |M|$. Any number of the lower values of mass $M_{c}$ can be found using
numerical methods. In accordance with the above, only the range $M_{c}>0.5$ is 
to be considered in numerical calculations. To find all points of the spectrum 
in the relevant ranges of $M_{c}$ and of the parameter $z$ for a fixed spin 
$l$, it suffices to restrict to seeking points $M_{c}$ at which the quantity 
$A_{0}$ in Eq. (24) vanishes but does not have a minimum or a maximum for some 
fixed value of $z$. In arbitrary small neighborhoods of such points in the mass
spectrum, the quantity $\chi_{lm}(l_{1})$ then obviously changes its sign for 
sufficiently large values of $l_{1}$. This plays the role of an algorithm for 
solving the mass problem numerically. Analyzing the dependence of any two 
chosen neighboring points of the mass spectrum on the parameter $z$, we can 
find whether a value $z_{0}$ exists such that in tending to it from one side, 
these points become arbitrarily close to each other but do not appear on the 
other side of $z_{0}$. If such a number $z_{0}$ exists, then the limit value of 
the two chosen points is the mass value $M_{c}$ at which the quantity $A_{0}$ 
has the zero value and an extremum.

In Figs. 1 and 2, in the cases corresponding to the respective relations (11) 
and (12), we show the dependence of the masses of the states with spins $1/2$, 
$3/2$, $5/2$, and $7/2$ on the parameter $z$.

\begin{figure}[ht]
\vspace{-1.2cm}
\centering \includegraphics[width=14.0cm]{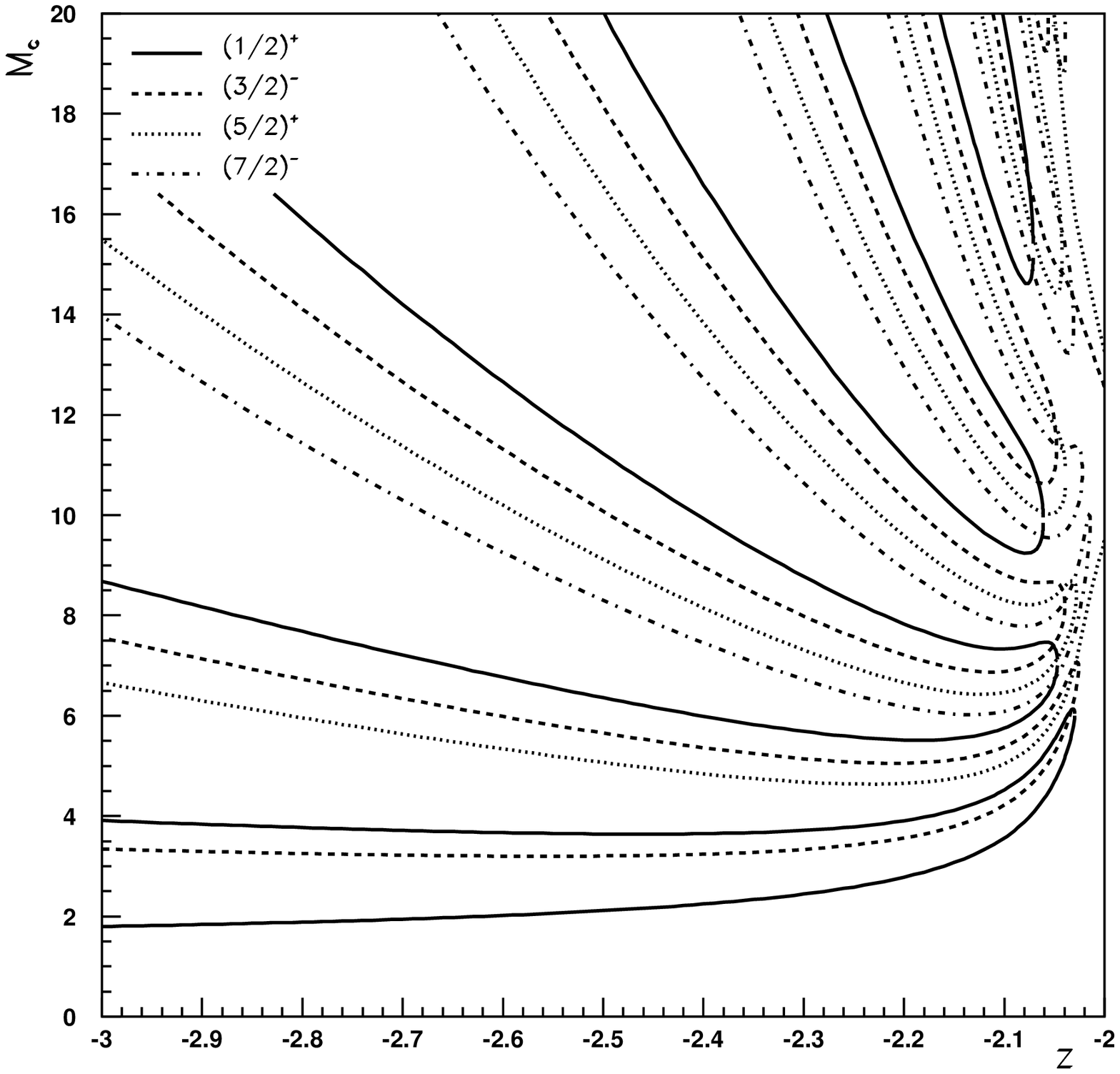}
\vspace*{-0.9cm}\\
\caption{Dependence of the mass levels on the parameter $z$ for $z < -2$ in the
theory with double symmetry generated by the axial four-vector representation 
of the orthochronous Lorentz group.}
\end{figure}

We first note several characteristics of the mass spectrum of the theory with 
double symmetry generated by the axial four-vector representation of the 
orthochronous Lorentz group (which corresponds to relation (12)). First, the 
mass spectrum is nonempty if the spatial parity of spin-$l$ particles is 
$(-1)^{l-1/2}$ and is empty if the parity is $(-1)^{l+1/2}$. Second, among the 
mass lines, there are pairs that terminate, merging at certain common limit 
points for $z_{0} < -2$; there are also single lines that exist in the entire 
range $z \in (-\infty,-2]$. Third, for all $z$, from the range $(-\infty,-2.1)$
at least, the levels of the mass spectrum have the same ordering in accordance 
with spin. It is the same as for $z=-2.645$ for example, where the lower masses
$M_{c}$ (with spin and parity $l^{P}$) are given by $1.982 (1/2)^{+}$, 
$3.209 (3/2)^{-}$, $3.687 (1/2)^{+}$, $5.470 (5/2)^{+}$, $6.143 (3/2)^{-}$, 
$6.964 (1/2)^{+}$, $9.709 (7/2)^{-}$, $10.72 (5/2)^{+}$, $11.91 (3/2)^{-}$, 
$13.35 (1/2)^{+}$, $17.82 (9/2)^{+}$, and $19.38 (7/2)^{-}$. The lowest-level 
masses with a given spin increase as the spin increases somewhat faster than 
the geometric progression. A sequence of mass levels taken in consecutive order
with the same spin is close to the geometric progression. Therefore, although 
the ratio of lowest-level masses with the spin and parity $(3/2)^{-}$ and 
$(1/2)^{+}$ in the above example is equal to the mass ratio of the $N(1520)$ 
resonance and the nucleon, the positions of levels with $l^{P} = (5/2)^{+}$, 
$(7/2)^{-}$, $(9/2)^{+}$, etc., are drastically different from the positions 
of the corresponding nucleonic resonances [15]. Fourth, for the parameter 
values close to $z=-2$, the level ordering in the mass spectrum in accordance 
with spin changes as $z$ changes. For example, for $z=-2$, the lower levels 
$M_{c}$ ($l^{P}$) are $9.506 (5/2)^{+}$, $12.14 (9/2)^{+}$, $12.55 (3/2)^{-}$, 
$13.25 (5/2)^{+}$, $15.77 (13/2)^{+}$, $16.03 (19/2)^{-}$, and 
$17.45 (21/2)^{+}$.

Among the mass-spectrum characteristics in the theory with double symmetry 
generated by the polar four-vector representation of the $L^{\uparrow}$ group 
(which corresponds to relation (11)), we note the following. First, particles 
at any value of spin have the same spatial parity $+1$. Second, the set of all 
mass lines is decomposed into pairs that merge and terminate at $z_{0} < -2$. 
Third, the lower-level masses with two consecutive spins, whenever they exist 
for a given $z$ in the range under consideration, differ from each other by an 
order of magnitude at least, which manifestly contradicts the baryon resonance 
picture.

\begin{figure}[ht]
\vspace{-1.2cm}
\centering \includegraphics[width=14.0cm]{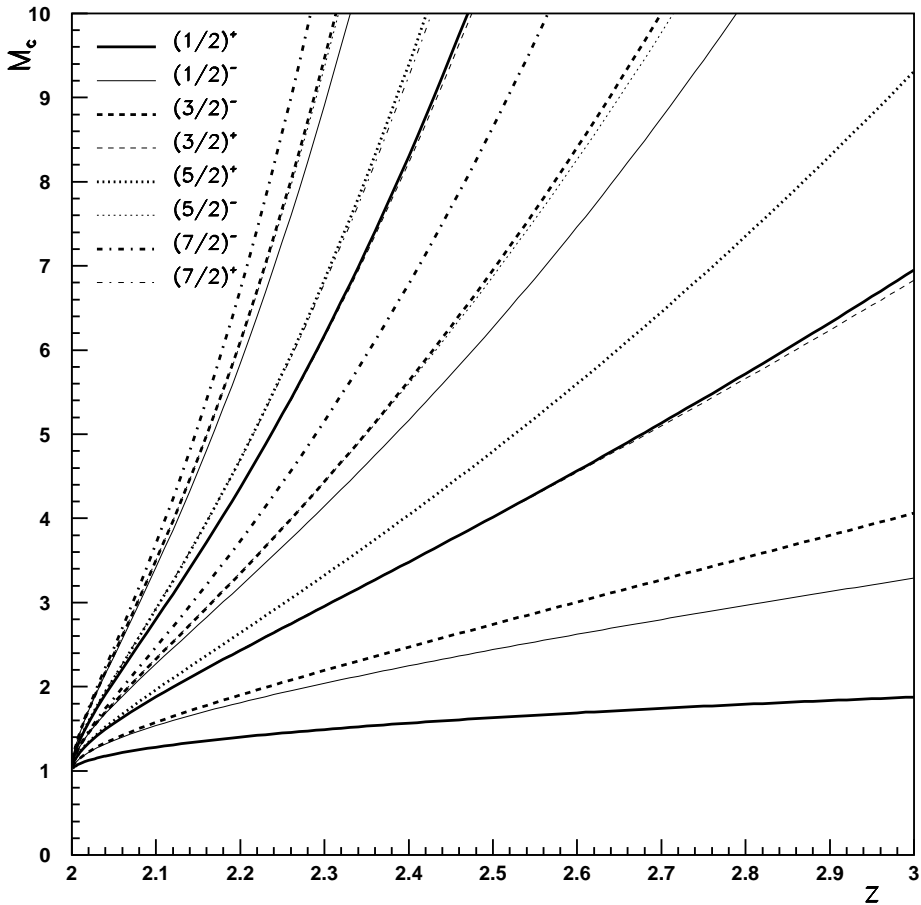}
\vspace*{-0.9cm}\\
\caption{Dependence of the mass levels on the parameter $z$ for $z > 2$ in the 
theory with double symmetry generated by the polar four-vector representation 
of the orthochronous Lorentz group.}
\end{figure}

\begin{center}
{\large \bf 5. Characteristics of the mass spectra in the parameter range 
$z \in (2, +\infty)$}
\end{center}

Because the inequalities $u > 1$ and $0 < w < 1$ are satisfied in the range 
$z \in (2, +\infty)$, formulas (23)--(25) hold, and the subsequent argument 
regarding the validity of condition (7) is applicable if $w$ is replaced with 
$u$ and $u$ with $w$ inthese formulas and in the corresponding argument. This 
implies the conclusion that for all values of $z \in (2, +\infty)$ in both 
versions of the theory corresponding to relations (11) and (12), the mass 
spectrum is discrete if it is nonempty.

It follows from relation (16) that in the range $z > 2$, the quantity 
$r(l_{1})/r(3/2)$ is positive and increases monotonically as $l_{1}$ increases.
This fact and Eq. (17) with $l_{1} > l$ lead to the inequality 
$|\chi_{lm} (l_{1})| > (1+1/(l_{1}-l))|\chi_{lm} (l)|$ for all $l$ and for all 
values of $M$ in the range $|M| \leq \mu_{1}/3$, where 
$\mu_{1} = |r(3/2)/c_{0}|$. In the range of $M$ specified, therefore, series 
(18) diverges, condition (7) is not satisfied, and mass-spectrum points are 
absent, i.e., the mass spectrum is bounded from below.

\begin{figure}[ht]
\vspace{-1.2cm}
\centering \includegraphics[width=14.0cm]{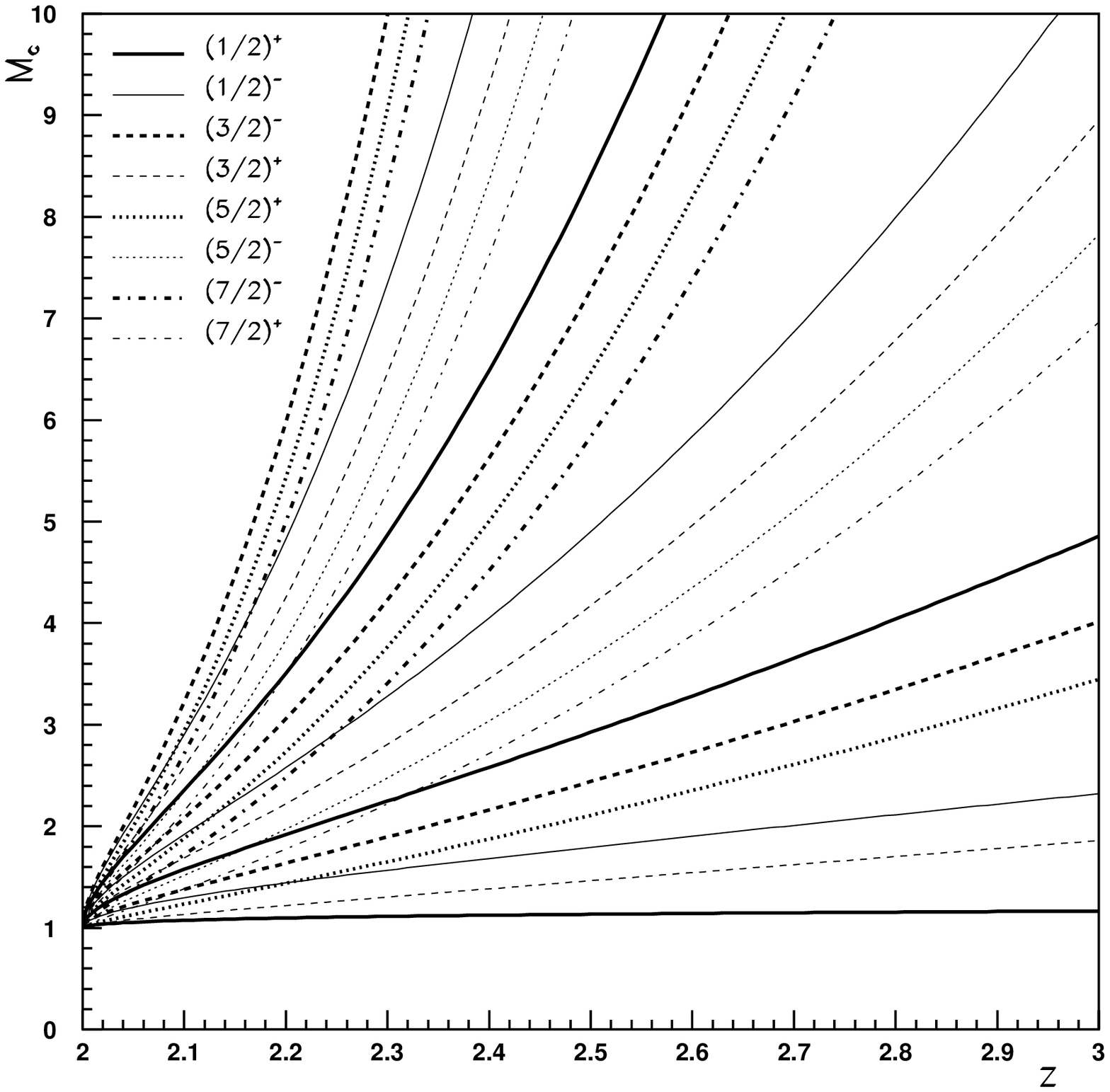}
\vspace*{-0.9cm}\\
\caption{Dependence of the mass levels on the parameter $z$ for $z > 2$ in the 
theory with double symmetry generated by the axial four-vector representation 
of the orthochronous Lorentz group.}
\end{figure}

In Figs. 3 and 4, corresponding to the respective formulas (11) and (12), we 
show the dependence of the lower mass levels on the parameter $z$ for spins 
$1/2$, $3/2$, $5/2$, and $7/2$, each of which is assigned the spatial parities 
$+1$ and $-1$. As $z \rightarrow 2+0$, the masses $M_{c}$ tend to unity on all 
lines in both versions of the theory. For any $z > 2$, the masses of several 
lower levels with consecutive spins $l$ and with the parities $(-1)^{l-1/2}$ 
approximate a geometric progression. This property also applies to the masses 
of several consecutive levels with the same spin and parity. This does not 
allow achieving any reasonable quantitative agreement between the lower levels 
of the theory with one parameter $z$ and the levels of nucleonic resonances. An 
essential difference between the two versions of the theory is in the order of 
relative positions of the lower levels with two consecutive spin values and a 
given parity. To illustrate all this, we give numerical examples below, where 
the mass ratio of the lower levels with $l^{P} = (3/2)^{-}$ and 
$l^{P} = (1/2)^{+}$ in the theory considered is the same as for the $N(1520)$ 
resonance and the nucleon.

We note several characteristic features of the mass spectrum for the version of
the theory with relation (11). First, the mass-level ordering in accordance 
with spin and parity is the same for all values of the parameter $z$ from the 
range $(2,+\infty )$. It is such as for $z=2.441$ for example, when the lowest 
masses $M_{c}$ and the corresponding spin and parity $l^{P}$ are given by 
$1.595(1/2)^{+}$, $2.328(1/2)^{-}$, $2.582(3/2)^{-}$, $3.699(1/2)^{+}$, 
$3.700(3/2)^{+}$, $4.346(5/2)^{+}$, $5.609(1/2)^{-}$, $6.111(5/2)^{-}$, 
$6.158(3/2)^{-}$, $7.522(7/2)^{-}$, $ 9.272(1/2)^{+}$, and $9.174(3/2)^{+}$. 
Second, certain mass levels with different $l^{P}$ differ from each other 
insignificantly in a considerable neighborhood of the point $z = 2$. Such a 
small difference occurs for the $(n+1)$th level with $l^{P} = (1/2)^{+}$ and 
the $n$th level with $l^{P} = (3/2)^{+}$, for the $(n+1)$th level with 
$l^{P} = (3/2)^{-}$ and the $n$th level with $l^{P} = (5/2)^{-}$, and for the 
$(n+1)$th level with $l^{P} = (5/2)^{+}$ and the nth level with 
$l^{P} = (7/2)^{+}$ ($n \geq 1$), etc.

For the version of the theory with relation (12), the general level ordering in
accordance with spin and parity can be different for different values of the 
parameter $z$. But the ordering of the lowest levels with some $l^{P}$ is the 
same for all $z > 2$. This ordering is as in the following example, where all
the levels with $M_{c} < 4$ are given, evaluated at $z = 2.261$: 
$1.108(1/2)^{+}$, $1.272(3/2)^{+}$, $1.521(1/2)^{-}$, $1.567(5/2)^{+}$, 
$1.794(3/2)^{-}$, $2.041(7/2)^{+}$, $2.119(1/2)^{+}$, $2.272(5/2)^{-}$, 
$2.573(3/2)^{+}$, $2.769(9/2)^{+}$, $2.999(1/2)^{-}$, $3.027(7/2)^{-}$, 
$3.342(5/2)^{+}$, $3.747(3/2)^{-}$, and $3.876(11/2)^{+}$.

On the whole, therefore, the doubly symmetric theory with one parameter $z$ 
connected with spontaneous secondary symmetry breaking gives the following 
qualitative picture of the mass spectra. The continuum part of the mass 
spectrum does not exist for any values of $z$. If the mass spectrum is nonempty 
for some $z$, then the mass levels are bounded from below by a positive 
quantity. In the parameter range $z > 2$, each value of spin and parity 
corresponds to a countable set of masses extending up to infinity, and the 
lowest mass level values with a given spin increase as the spin increases. The 
theory with one parameter $z$ does not give a satisfactory quantitative 
agreement with the nucleon resonance levels.

\begin{center}
{\large \bf 6. Comparison of the mass spectrum of the theory with two 
parameters $z_{i}$ with nucleon resonance levels}
\end{center}

The general situation with the mass spectra of the theory involving two bosonic
fields of the ISFIR class with nonvanishing vacuum expectations of their scalar
components (scalar with respect to the group $L^{\uparrow}$) is quite rich with
different versions. In what follows, we consider only one version in some 
detail with the aim of a tentative quantitative comparison of the theoretical 
mass spectrum with the nucleon resonance levels.

We consider the theory with the double symmetry generated by the polar 
four-vector representation of the orthochronous Lorentz group (with the 
corresponding relation (11)). With two parameters $z_{i}$, the quantity 
$r(l_{1})$ in Eq. (17) can be written as
\begin{equation}
r(l_{1}) = f_{0} \left( \frac{q_{1}(\frac{1}{2}, l_{1})}{q_{10}} 
+ f_{2} \frac{q_{2}(\frac{1}{2}, l_{1})}{q_{20}} \right) ,
\end{equation}
where $q_{i}(1/2, l_{1})$ depends on $z_{i}$ and is given by formula (16). It 
is obvious that in addition to the normalization constant $f_{0}/c_{0}$, this 
theory involves three free parameters: $z_{1}$, $z_{2}$, and $f_{2}$. We choose 
the following restrictions on them: $z_{1} > 2$, $|z_{2}| < 2$, and 
$|f_{2}| < 1$. In this parameter range, as $l_{1}$ increases, the quantity 
$|q_{1}(1/2, l_{1})|$ increases monotonically and $q_{2}(1/2, l_{1})$ 
oscillates with a decreasing amplitude. The quantities $r(l_{1})$ and 
$f_{0} q_{1}(1/2, l_{1})/q_{10}$ have the same sign for all $l_{1}$ and can be 
significantly different from each other only for small values of $l_{1}$. The 
parameter $z_{2}$ has no effect on the asymptotic behavior of the quantities 
$r(l_{1})$ and $\chi_{lm} (l_{1})$ as $l_{1} \rightarrow +\infty$. Formulas 
(23)--(25) are valid if $w$ in them is replaced with $u_{1}$; the conclusions 
regarding the mass spectra following from these formulas are also valid. 
Therefore, in the considered version of the theory with two parameters $z_{1}$ 
and $z_{2}$, compared with the version of the theory with a single parameter 
$z_{1}$, only mass spectrum levels with the lowest values of the spin $l$ can 
significantly change their positions.

We now note a very important circumstance inherent in the theory of 
infinite-component fields considered. Let the ground fermionic and the ground 
bosonic levels of the theory correspond to the respective particles $F$ with 
spin $1/2$ and $B$ with spin $0$, and let an excited fermionic level exist in 
the theory with the corresponding particle $F^{*}$ of spin $l$. We consider the 
amplitude ${\cal M}$ of the decay $F^{*} \rightarrow FB$ in the rest frame of 
the resonance $F^{*}$ described by Lagrangian (4). Because the particles $F$ 
and $B$ have nonzero velocities in this reference frame, the corresponding
infinite-component fields of the ISFIR class have nonzero components with the 
respective all half-integer and all integer spins. Therefore, the amplitude
${\cal M}$ has nonzero terms in which the components of the $F^{*}$, $F$, and 
$B$ fields are described by the following sets of spins 
$\{ l_{F^{*}}, l_{F}, l_{B} \}$: $\{ l,l,0 \}$, $\{ l,l-1,1 \}$, $\{ l,l,1 \}$, 
$\{ l,l+1,1 \}$, $\{ l,l-2,2 \}$, $\ldots$, $\{ l,l+2,2 \}$, $\ldots$. The 
contribution of a given spin $l'$ of the fermion $F$ and spin $l''$ of the 
boson $B$ to the amplitude ${\cal M}$ can be estimated only if the state 
vectors of the particles $F^{*}$, $F$, and $B$ are known in their rest frames. 
Because experimental conclusions about a resonance spin are obtained from the 
partial-wave analysis, based on the representations of the three-dimensional 
rotation group but not on representations of the Lorentz group, it follows that
the aforesaid may be manifested in the experimentalist opinion regarding the 
existence of a group of several resonances that have the same mass and differ 
from each other only by their spins. Therefore, in comparing the theoretical 
mass spectrum with the experimental one, we must pay special attention to such 
groups of resonances and make the decision regarding the number of the 
corresponding theoretical levels.

The simplest proposed correspondence between the theory version under 
consideration and the experimental picture of nucleon resonances is given in 
Table 1. It corresponds to the normalization constant and the free parameters 
chosen as $f_{0}/c_{0} = -939/2.4686$ MeV, $z_{1} = 2.036$, $z_{2} = 0.14$, 
and $f_{2} = -0.6724$. The parameters $z_{2}$ and $f_{2}$ are chosen such that 
the lowest-level masses with $l^{P} = (3/2)^{-}$ and $l^{P} = (5/2)^{+}$ are 
respectively equal to 1508 and 1675 MeV. The parameter $z_{1} = 2.036$ 
determines particle states with large spins. As the experimental masses of 
almost all resonances, we take their pole positions. The Breit--Wigner masses 
(indicated with the BW superscript in the table) are taken, first, for those 
few resonances whose pole positions are not given in [15] and, second, for the 
Roper resonance $N(1440)$ because two poles, $1370-114i$ and $1360-120i$, have 
been found in the neighborhood of 1440 MeV with their parameters strongly 
different from the usual $M = 1470$ MeV and $\Gamma = 545$ MeV [16].

A tentative comparison of the mass spectra of the theory considered here with 
levels of nonnucleon resonances has not been made yet. In particular, it 
requires the singlet--octet separation of the $\Lambda$ resonances and an 
octet--decuplet separation of the $\Sigma$ and $\Xi$ resonances in accordance 
with the internal symmetry group $SU(3)$. The description of the $\Delta$
resonances is supposedly to be given by the $S^{5/2}$ representation of the 
$L^{\uparrow}_{+}$ group described by formula (11) with $k_{1} = 5/2$. In
this case, the lowest mass level has the spin $3/2$. Analyzing mass spectra of 
bosonic fields of the ISFIR class in any versions of the doubly symmetric 
theory will become possible only after finding the structure of the Lagrangian 
of the four-particle interaction of such fields with each other and the ensuing 
spontaneous violation of secondary symmetry.

\begin{center}
\begin{tabular}[ht]{|cc|cccc|}
\multicolumn{6}{r}{\bf Table 1} \\
\hline
\multicolumn{2}{|c|}{Theory} 
&\multicolumn{4}{c|}{Experiments [15]} \\
\hline
$l^{P}$ 
&\begin{tabular}{c} Mass \\ (MeV) \\ \end{tabular}
&$l^{P}$ &Resonance 
&\begin{tabular}{c} Mass \\ (MeV) \\ \end{tabular} 
&Status \\
\hline
\hspace{0.14 cm} $\frac{1}{2}^{+}$ &939   &\hspace{0.14 cm} $\frac{1}{2}^{+}$ 
&$N$       &939       &**** \\
\hspace{0.14 cm} $\frac{1}{2}^{+}$ &1481  &\hspace{0.14 cm} $\frac{1}{2}^{+}$ 
&$N(1440)$ &1430-1470$^{\rm BW}$ &**** \\
\hspace{0.14 cm} $\frac{1}{2}^{-}$ &1487  &\hspace{0.14 cm} $\frac{1}{2}^{-}$ 
&$N(1535)$ &1495-1515 &**** \\
\hspace{0.14 cm} $\frac{3}{2}^{-}$ &1508  &\hspace{0.14 cm} $\frac{3}{2}^{-}$ 
&$N(1520)$ &1505-1515 &**** \\
\hspace{0.14 cm} $\frac{3}{2}^{+}$ &1661  
&$\left\{ \begin{array}{l} \frac{1}{2}^{-} \\ \frac{5}{2}^{-} \\ 
\frac{3}{2}^{+} \\ \end{array} \right.$
&\begin{tabular}{c} $N(1650)$ \\ $N(1675)$ \\  $N(1720)$ \\ \end{tabular}
&\begin{tabular}{c} 1640-1680 \\ 1655-1665 \\  1650-1750 \\ \end{tabular}
&\begin{tabular}{c} **** \\ {*}*** \\ {*}*** \\ \end{tabular} \\
\hspace{0.14 cm} $\frac{5}{2}^{+}$ &1675  
&$\left\{ \begin{array}{l} \frac{5}{2}^{+} \\ 
\frac{3}{2}^{-} \\ \frac{1}{2}^{+} \\ \end{array} \right.$
&\begin{tabular}{c} $N(1680)$ \\ $N(1700)$ \\ $N(1710)$ \\ \end{tabular}
&\begin{tabular}{c} 1665-1675 \\ 1630-1730 \\ 1670-1770 \\ \end{tabular}
&\begin{tabular}{c}  **** \\ {*}** \\ {*}** \\ \end{tabular} \\
$\left\{ \begin{array}{l} \frac{5}{2}^{-} \\ \frac{1}{2}^{-} 
\\ \end{array} \right.$
&\begin{tabular}{c} 1892 \\ 1923 \\ \end{tabular}
&$\left\{ \begin{array}{l} \frac{3}{2}^{+} \\ \frac{7}{2}^{+} \\ 
\end{array} \right.$
&\begin{tabular}{c} $N(1900)$ \\ $N(1990)$ \\ \end{tabular}
&\begin{tabular}{c} $\approx 1900^{\rm BW}$ \\ 1870-1930 \\ 
\end{tabular}
&\begin{tabular}{c} ** \\  {*}* \\ \end{tabular} \\
\hspace{0.14 cm} $\frac{7}{2}^{-}$ &1940 &\hspace{0.14 cm} $\frac{7}{2}^{-}$ 
&$N(2190)$ &1950-2150 &**** \\
$\left\{ \begin{array}{l} \frac{3}{2}^{-} \\ \frac{1}{2}^{+} \\ 
\frac{5}{2}^{+} \\ \end{array} \right.$
&\begin{tabular}{c} 1995 \\ 2004 \\ 2144 \\ \end{tabular}
&$\left\{ \begin{array}{l} \frac{5}{2}^{+} \\  \frac{3}{2}^{-} \\
\frac{1}{2}^{+} \\ \end{array} \right.$
&\begin{tabular}{c} $N(2000)$ \\ $N(2080)$ \\ $N(2100)$ \\ \end{tabular}
&\begin{tabular}{c} $\approx 2000^{\rm BW}$ \\ 1980-2120 \\ 
2080-2160 \\ \end{tabular}
&\begin{tabular}{c} ** \\ {*}* \\ {*} \\ \end{tabular} \\
\hspace{0.14 cm} $\frac{3}{2}^{+}$ & 2140 &\hspace{0.14 cm} $\frac{1}{2}^{-}$ 
&$N(2090)$ &2080-2220 &* \\ \hspace{0.14 cm} $\frac{7}{2}^{+}$ &2179 
&$\left\{ \begin{array}{l} \frac{5}{2}^{-} \\ \frac{9}{2}^{-} \\
\end{array} \right.$
&\begin{tabular}{c} $N(2200)$ \\ $N(2250)$ \\ \end{tabular}
&\begin{tabular}{c} 2040-2160 \\ 2080-2200 \\ \end{tabular}
&\begin{tabular}{c} ** \\ {*}*** \\ \end{tabular} \\
\hspace{0.14 cm} $\frac{9}{2}^{+}$ &2244 &\hspace{0.14 cm} $\frac{9}{2}^{+}$ 
&$N(2220)$ &2100-2240 &**** \\
$\ldots$ &$\ldots$ & & & & \\
\hspace{0.14 cm} $\frac{11}{2}^{-}$ &2547 &\hspace{0.14 cm} $\frac{11}{2}^{-}$ 
&$N(2600)$ &2550-2750$^{\rm BW}$ &*** \\
$\ldots$ &$\ldots$ & & & & \\
\hspace{0.14 cm} $\frac{13}{2}^{+}$ &2919 &\hspace{0.14 cm} $\frac{13}{2}^{+}$ 
&$N(2700)$ &$\approx 2700^{\rm BW}$ &* \\
\hline
\end{tabular}
\end{center}

{\bf Acknowledgments.} The author is very grateful to E.E.Boos, V.I.Savrin, 
and N.P.Yudin for the useful discussions and the support of his work.

\end{small}
\end{document}